\documentclass[aip,graphicx]{revtex4-1}
\usepackage{graphicx}
\usepackage{bm}
\usepackage{dcolumn}

\begin{document}

\title{Train-Resolved Statistical Recovery of Weak SAXS Signals in Liquids at the European XFEL}

\author{Carles Serrat}
\email{carles.serrat-jurado@upc.edu}
\affiliation{Departament de F\'isica, Universitat Polit\`ecnica de Catalunya--BarcelonaTech (UPC), Terrassa, Barcelona, Spain}

\author{Asier Garc\'ia}
\affiliation{IMDEA Nanociencia, Madrid, Spain}

\author{Biel Serrat}
\affiliation{Escola d'Enginyeria de Telecomunicaci\'o i Aeroespacial de Castelldefels, Universitat Polit\`ecnica de Catalunya--BarcelonaTech (UPC), Castelldefels, Barcelona, Spain}

\author{Angelo Beratto-Ramos}
\affiliation{IMDEA Nanociencia, Madrid, Spain}

\author{Johan Bielecki}
\affiliation{European XFEL, Schenefeld, Germany}

\author{Huijong Han}
\affiliation{European XFEL, Schenefeld, Germany}

\author{Sara Hern\'andez}
\affiliation{IMDEA Nanociencia, Madrid, Spain}

\author{Tokushi Sato}
\affiliation{European XFEL, Schenefeld, Germany}

\author{Joana Valerio}
\affiliation{European XFEL, Schenefeld, Germany}

\author{Mohammad Vakili}
\affiliation{European XFEL, Schenefeld, Germany}

\author{Egor Sobolev}
\affiliation{European XFEL, Schenefeld, Germany}

\author{Katerina Doerner}
\affiliation{European XFEL, Schenefeld, Germany}

\author{Chan Kim}
\affiliation{European XFEL, Schenefeld, Germany}

\author{Majed Chergui}
\affiliation{Lausanne Centre for Ultrafast Science (LACUS), ISIC, Ecole Polytechnique F\'ed\'erale de Lausanne, Lausanne, Switzerland}
\affiliation{Elettra Sincrotrone Trieste S.C.p.A., Trieste, Italy}

\begin{abstract}
We present a train-resolved SAXS methodology for recovering weak scattering signals from high-repetition-rate XFEL datasets and apply it to aqueous L-cysteine solutions measured at the European XFEL. For each XFEL train, an energy-normalized ON--OFF radial profile was reconstructed. Matched cysteine and water train profiles were then compared using an independent scale-plus-offset model, and the resulting residuals were corrected using transmission-matched water--water controls acquired under comparable conditions.
The 0.5 M dataset reveals a reproducible residual SAXS signal that persists after removal of both the fitted scale-plus-offset contribution and the corresponding water controls. The residual exhibits a characteristic sign-changing dependence on momentum transfer, with a positive contribution over approximately \(0.05<q<0.15~\mathrm{\AA^{-1}}\) followed by a negative contribution over approximately \(0.16<q<0.30~\mathrm{\AA^{-1}}\). The signal is weak at the lowest transmission (\(T=0.0099\)) and becomes stronger at higher transmissions (\(T=0.21\) and \(T=0.32\)), consistent with increasing incident XFEL fluence.
Convergence and block-averaging analyses show that the residual emerges as independent train pairs are accumulated and that the integrated residual observables exhibit variability consistent with \(1/\sqrt{N}\), where \(N\) is the block size. The observed structure therefore represents a statistically robust ensemble property of the dataset rather than a small number of anomalous events.
These measurements establish a reproducible transmission-dependent residual SAXS contribution in aqueous L-cysteine solutions that is not captured by detector-wide scaling or additive corrections and exceeds the residual structure observed in matched water controls. 
While the microscopic origin remains unresolved, the results demonstrate that suitable train-level ON--OFF observables combined with matched controls can recover weak SAXS signals with uncertainty scaling close to the expected \(1/\sqrt{N}\) behaviour in high-repetition-rate XFEL experiments.
\end{abstract}

\pacs{}

\maketitle

\section{Introduction}

Small-angle X-ray scattering (SAXS) has become a powerful tool for probing structure and structural dynamics in solution across length scales ranging from nanometre-scale correlations to mesoscale heterogeneity \cite{Graewert2020SAXSModeling}. The advent of X-ray free-electron lasers (XFELs) has extended these capabilities into regimes of extreme photon flux and ultrafast temporal resolution, enabling SAXS measurements under conditions that are inaccessible at conventional synchrotron sources \cite{Blanchet2023XFELSAS}.

Recent developments at the European XFEL and other XFEL facilities have demonstrated the feasibility of X-ray scattering measurements using liquid jets and fast area detectors \cite{Blanchet2023XFELSAS,Allahgholi2019AGIPDMegahertz}. In particular, the European XFEL enables train-resolved acquisition schemes in combination with megahertz-rate detectors, in which scattering data are analyzed separately for individual XFEL pulse trains before ensemble averaging. These advances provide new opportunities for investigating biological and chemical systems under intense irradiation conditions. One of the principal motivations for high-repetition-rate XFEL operation is the possibility of detecting extremely weak signals through statistical averaging over large ensembles of measurements. In the ideal case of statistically independent observations, the associated uncertainty is expected to decrease as \(1/\sqrt{N}\), where \(N\) is the number of measurements. In practice, however, the achievable sensitivity is often limited by detector-wide fluctuations, normalization uncertainties, beam instabilities, liquid-jet variability, and other correlated contributions that do not necessarily follow ideal statistical scaling.

At the same time, high-repetition-rate operation introduces significant challenges. The measured scattering signal may contain contributions from equilibrium sample structure, beam-induced modifications of the liquid, detector and normalization effects, jet instabilities, and pulse-to-pulse or train-to-train fluctuations. When the effect of interest is weak, separating genuine sample-dependent scattering from these competing contributions becomes non-trivial. An important question is therefore whether train-resolved observables can recover the expected statistical averaging behaviour once dominant common-mode contributions are removed. Establishing such behaviour is essential for assessing the feasibility of future XFEL experiments targeting extremely weak fluence-dependent or nonlinear signals.

In conventional SAXS difference analysis, sample and reference profiles are typically normalized and averaged before comparison. While effective for strong and reproducible signals, this approach can obscure weak scattering contributions when larger multiplicative and additive variations are present between measurements. Such variations may arise from fluctuations in pulse energy, detector response, liquid-jet conditions, beam-position drifts, masking effects, or other sources of experimental variability. Consequently, the reliable identification of weak fluence-dependent SAXS signatures requires carefully matched controls acquired under equivalent experimental conditions.

The importance of such controls has been emphasized previously in XFEL-SAXS studies, where subtraction accuracy and experimental reproducibility often determine the ultimate sensitivity of the measurement \cite{Blanchet2023XFELSAS}. Residual differences remaining after conventional normalization procedures may therefore contain either genuine sample-dependent information or artefacts arising from imperfect experimental matching. Distinguishing between these possibilities requires explicit evaluation of the residual structure itself.

In the present work, we analyze SAXS measurements of aqueous L-cysteine solutions acquired at the European XFEL using the SPB/SFX instrument and the Adaptive Gain Integrating Pixel Detector (AGIPD) system \cite{Mancuso2019SPBSFX,Allahgholi2019AGIPD}. Our objective is to determine whether the difference between cysteine and water scattering can be fully described by a global scale-plus-offset transformation, or whether a reproducible momentum-transfer-dependent residual remains after removal of these dominant components.

The analysis focuses on the 0.5 M cysteine dataset, for which transmission-matched water references and water--water controls were acquired under closely comparable experimental conditions. We show that a reproducible residual SAXS signal remains after subtraction of both the scale-plus-offset contribution and the matched water controls. The residual exhibits a characteristic sign-changing dependence on momentum transfer and increases systematically with transmission. Statistical analyses further demonstrate that the signal represents a reproducible ensemble property of the train-pair dataset.

The present work is intentionally conservative in its interpretation. Our objective is not to assign a specific microscopic mechanism to the observed residual signal, but rather to establish its existence, quantify its statistical robustness, and assess its dependence on the incident XFEL fluence. The resulting observations provide experimental evidence for a reproducible transmission-dependent residual SAXS contribution in aqueous cysteine solutions that survives both scale-plus-offset correction and matched water controls.

\section{Experimental Methods}

\subsection{European XFEL measurements}

The SAXS measurements analyzed in this work were acquired at the European X-ray Free-Electron Laser (EuXFEL) using the SPB/SFX instrument \cite{Mancuso2019SPBSFX}. X-ray pulses with a photon energy of 6 keV and a pulse duration of approximately 25 fs were focused onto a continuously refreshed liquid jet. The X-ray beam was focused to an approximately \(5\times25~\mu\mathrm{m}^2\) spot at the interaction region, and data were acquired using the standard EuXFEL pulse-train structure. Scattered photons were recorded using the AGIPD positioned at a sample--detector distance of approximately 1.5 m, enabling train-resolved SAXS measurements over a broad range of momentum transfer \cite{Allahgholi2019AGIPD}.
Samples were delivered using a liquid jet with a diameter of approximately \(8~\mu\mathrm{m}\) and a jet velocity of approximately \(30~\mathrm{m\,s^{-1}}\) \cite{Vakili2022SampleDelivery}.

The experiment was designed to investigate the response of aqueous L-cysteine solutions to intense XFEL irradiation. Particular emphasis was placed on identifying weak fluence-dependent scattering signatures that could remain hidden beneath substantially larger global intensity variations.

Additional beamline and detector parameters will be reported elsewhere and are not required for the methodological analysis presented here.

\subsection{Samples and transmission settings}

Two aqueous L-cysteine concentrations were measured, \(c=0.5~\mathrm{M}\) and \(c=1.0~\mathrm{M}\), together with pure water reference measurements. The L-cysteine solutions were prepared in 1 mM HCl (\(pH\approx5\)) following standard procedures and used under the same conditions as the corresponding water references.

The primary analysis presented in this article focuses on the 0.5 M dataset because it contains the most complete set of transmission-matched water references and water--water controls. These controls are essential for distinguishing sample-dependent residual scattering from run-to-run experimental variability.

Measurements were acquired under several attenuation conditions defined by the beamline filter settings. Throughout this article, datasets are labelled by the corresponding transmission value \(T\). Higher transmission corresponds to higher incident XFEL fluence. The principal transmission settings used in the analysis were \(T=0.0099\), \(T=0.21\), and \(T=0.32\). For each transmission, cysteine and water measurements were acquired in close temporal proximity in order to minimize the influence of slow experimental drifts.

\subsection{Run organization and control strategy}

The run combinations used in the primary analysis are summarized in Table~\ref{tab:run_overview}.

\begin{table}[htbp]
\centering
\caption{Run pairs used in the primary 0.5 M analysis.}
\label{tab:run_overview}
\begin{tabular}{ccc}
\hline
Transmission & Cysteine runs & Water runs \\
\hline
0.0099 & 28--29 & 26--27 \\
0.21   & 24--25 & 22--23 \\
0.32   & 18--19 & 20--21 \\
\hline
\end{tabular}
\end{table}

In addition to the cysteine--water comparisons, independent water--water controls were constructed from the same water datasets: run 27 versus run 26, run 23 versus run 22, and run 21 versus run 20.

These controls play a central role in the analysis because they provide a direct measurement of the residual structure expected from detector reproducibility limits, beam fluctuations, normalization imperfections, liquid-jet variability, and other run-dependent effects in the absence of a sample change.

The scientific conclusions of this work are therefore based not only on cysteine--water comparisons but also on explicit evaluation of the corresponding water--water residuals under matched experimental conditions.

\begin{figure*}[t]
\centering
\includegraphics[width=\textwidth]{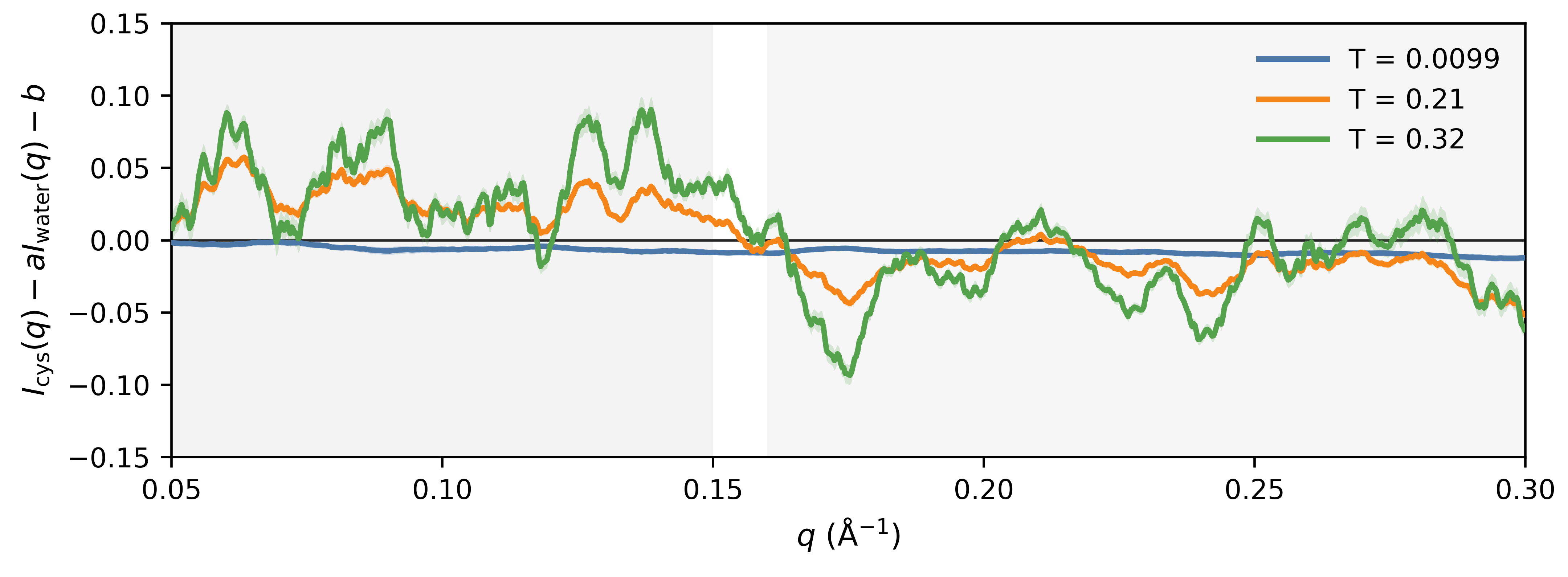}
\caption{
Train-averaged residual structure for the 0.5 M cysteine--water comparisons after independent train-by-train scale-plus-offset subtraction. The shaded regions indicate the momentum-transfer intervals used to define the integrated residual observables \(A_+\) and \(A_-\); the unshaded interval between them is excluded from the integration. The residual amplitude increases strongly with transmission, while the two highest-transmission datasets exhibit a similar momentum-transfer dependence. Curves are lightly smoothed for visual clarity; all quantitative analyses were performed using the original unsmoothed data.
}
\label{fig:deep0p5M_residuals_raw}
\end{figure*}

\begin{figure*}[t]
\centering
\includegraphics[width=\textwidth]{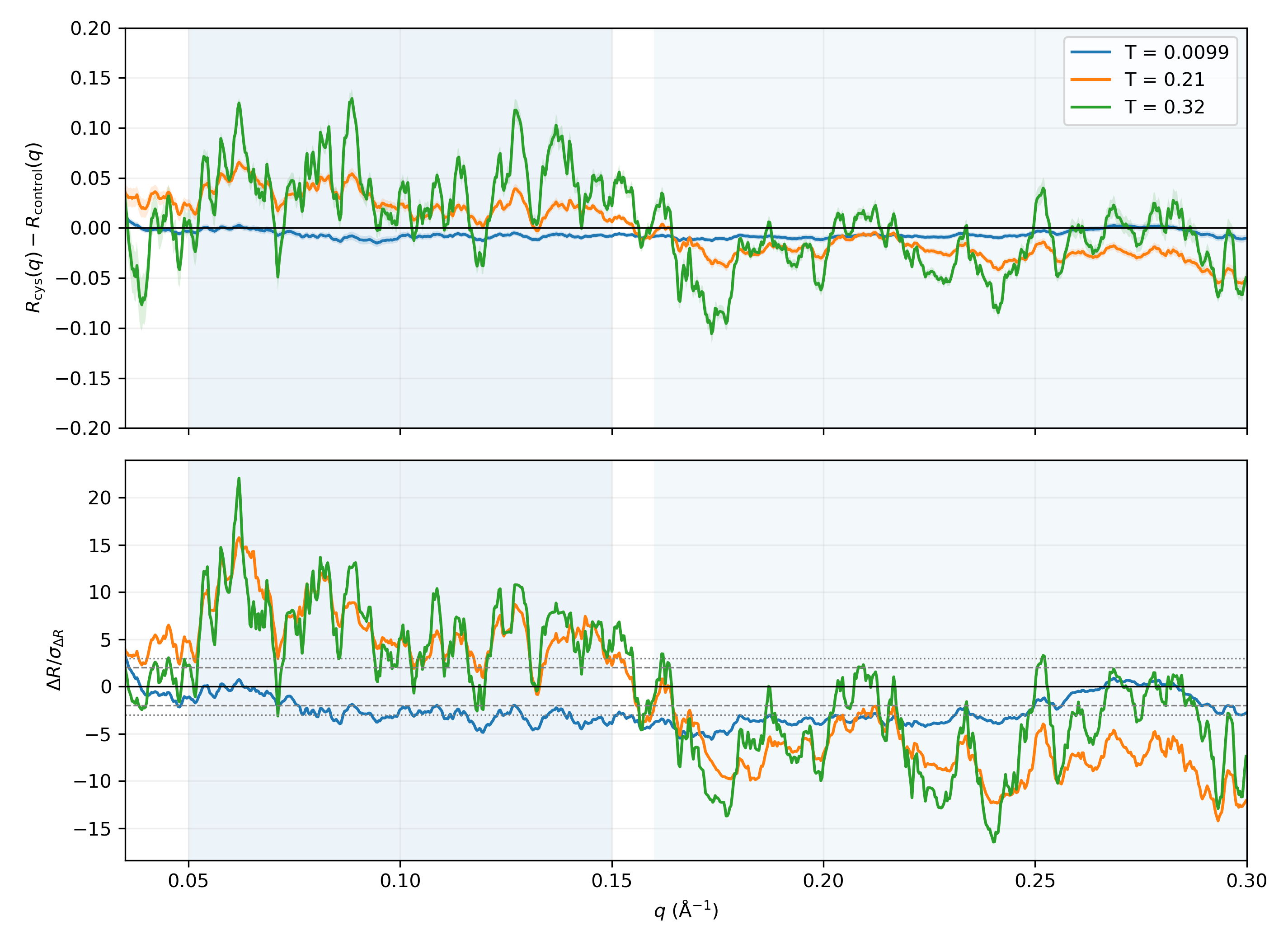}
\caption{
Control-subtracted residual SAXS structure for the 0.5 M cysteine dataset. The upper panel shows the control-corrected residual, \(\Delta R(q)\), obtained by subtracting the corresponding transmission-matched water--water control from each cysteine--water residual. The lower panel shows the associated significance, \(\Delta R(q)/\sigma_{\Delta R}(q)\). Shaded regions indicate the momentum-transfer intervals used to define the integrated residual observables \(A_+\) and \(A_-\), while the intermediate unshaded region is excluded from the integration. The two highest-transmission datasets exhibit closely similar residual profiles with substantially larger amplitudes than the lowest-transmission dataset, indicating a reproducible transmission-dependent signal. Curves are lightly smoothed for visual clarity; all quantitative analyses were performed using the original unsmoothed data.}
\label{fig:deep0p5M_cys_minus_control}
\end{figure*}

\section{Train-resolved scale-plus-offset analysis}

The objective of the analysis is to determine whether the difference between cysteine and water scattering can be explained entirely by global intensity variations, or whether a reproducible momentum-transfer-dependent residual remains after removal of these dominant contributions.

All analysis was performed at the train level. For each run, ON--OFF radial profiles were computed independently for individual XFEL trains. This procedure removes the dominant static background contribution while preserving train-to-train fluctuations and possible sample-dependent scattering differences.

For a given transmission setting, train-resolved cysteine profiles were paired with train-resolved water profiles acquired under matched experimental conditions. The comparison was performed independently for each train pair rather than after averaging over the complete dataset.

The cysteine profile was modelled as a scaled and offset version of the corresponding water profile,

\[
I_{\rm cys}(q)=a\,I_{\rm water}(q)+b,
\]

where \(a\) accounts for multiplicative intensity differences and \(b\) represents a detector-wide additive contribution. Both parameters were determined independently for each train pair using least-squares minimization over the selected fitting range.

The residual profile was then defined as

\[
R(q)=I_{\rm cys}(q)-a\,I_{\rm water}(q)-b.
\]

If the difference between cysteine and water scattering is fully described by global intensity scaling and additive offsets, the residual should fluctuate around zero within statistical uncertainty. Conversely, any reproducible structure remaining in \(R(q)\) indicates a scattering contribution that cannot be absorbed into the fitted scale and offset parameters.

Fitting averaged profiles is not generally equivalent to averaging residuals obtained from independently fitted train pairs. In the present implementation, the fit was performed independently for every train pair, making the residual distribution directly accessible for subsequent statistical analysis. 

To evaluate the level of residual structure expected from experimental reproducibility limits, the same procedure was applied to transmission-matched water--water comparisons. These controls do not represent a sample signal. Instead, they provide an empirical estimate of the residual structure that survives the scale-plus-offset correction in the absence of any sample change, including contributions from run-to-run variations, detector effects, normalization imperfections, liquid-jet variability, and other experimental common modes.

The control-corrected residual,

\[
\Delta R(q)=
R_{\rm cys-water}(q)
-
R_{\rm water-water}(q),
\]

represents the central observable of the present work. By construction, it removes the fitted scale-plus-offset contribution and corrects for the mean residual structure measured in the corresponding water-only controls. The remaining signal can therefore be evaluated relative to the residual structure observed when identical samples are compared under matched experimental conditions.

For each transmission, train-averaged residual profiles were computed from 200 uniformly sampled train pairs. The same number of train pairs was used for all transmission settings to ensure comparable statistical precision across the datasets. Statistical robustness was subsequently assessed through convergence analysis and block-averaging statistics performed on the train-pair ensemble. Convergence tests performed using subsets containing 50, 100, and 200 train pairs showed progressive stabilization of the residual profiles and integrated observables over the investigated range. A more extensive convergence analysis using larger train ensembles is currently in progress.

\section{Dataset selection and experimental controls}
\label{sec:dataset_assessment}

The interpretation of weak residual SAXS signals requires a dataset in which sample-dependent differences can be separated from run-to-run variations. For this reason, the quantitative analysis in this work is restricted to measurements that satisfy three criteria: cysteine and water data were acquired at the same transmission, the measurements were obtained in close temporal proximity, and an independent water--water comparison was available at the same transmission.

The 0.5 M cysteine dataset at \(T=0.0099\), \(T=0.21\), and \(T=0.32\) satisfies these requirements. The corresponding cysteine--water comparisons are listed in Table~\ref{tab:run_overview}, and the water--water controls were constructed from the same water runs. These controls provide an internal measure of the residual structure expected from detector reproducibility, beam fluctuations, normalization imperfections, liquid-jet variability, and other non-sample-related effects.

As an additional methodological validation, azimuthal integration was verified by comparing two independent processing routes: averaging individual integrated profiles and integrating averaged detector images. The resulting differences were negligible compared with the residual signals discussed below, indicating that the observed structures are not sensitive to the integration procedure.

The intermediate transmission point \(T=0.046\) was also examined during the analysis. However, the acquisition sequence at this transmission was less controlled. The water measurement was obtained in run 14, while run 15, intended as the corresponding cysteine run, was interrupted when the sample reservoir was exhausted. The subsequent cysteine measurements, runs 16 and 17, were acquired only after replacement of the sample. Although run 14 can be used as a water reference for runs 16 and 17, this sequence does not provide the same experimental symmetry as the primary transmission-matched datasets.
The resulting residual does not follow the trend defined by the three controlled transmission settings, and it cannot be determined from the present data whether this deviation is physical or due to the modified acquisition conditions. The \(T=0.046\) measurements are therefore excluded from the primary fluence-dependent analysis.

The 1 M cysteine measurements were also inspected using the same train-resolved scale-plus-offset procedure. Residual structures are observed in these data, but the dataset does not contain a complete set of transmission-matched water controls acquired under comparable conditions. Consequently, the 1 M residuals cannot be unambiguously separated from run-dependent variations. These measurements are therefore not used to establish concentration dependence or quantitative fluence scaling in the present work.

The primary dataset used for the scientific conclusions of this article is therefore restricted to the 0.5 M cysteine measurements at \(T=0.0099\), \(T=0.21\), and \(T=0.32\), together with their corresponding water references and water--water controls. This restriction is conservative but essential: it ensures that the residual SAXS signal discussed below is evaluated only where the experimental controls are sufficient to distinguish sample-dependent structure from reproducibility limits.

\section{Observation of residual SAXS structure in 0.5 M cysteine}

The primary objective of the present work is to determine whether any reproducible SAXS structure remains after removal of the dominant multiplicative and additive differences between cysteine and water measurements. The analysis therefore proceeds in two stages. First, residuals are examined after train-resolved scale-plus-offset subtraction. Second, the corresponding water--water control residuals are removed to isolate the component that exceeds the experimental reproducibility limits measured within the water dataset itself.

\subsection{Residuals after scale-plus-offset subtraction}

Figure~\ref{fig:deep0p5M_residuals_raw} shows the train-averaged residuals obtained after independent train-by-train scale-plus-offset subtraction for the three principal transmission settings.

The residuals are not flat in momentum transfer. In the lowest-transmission dataset (\(T=0.0099\)), the residual remains relatively weak over the full \(q\) range. In contrast, the higher-transmission datasets (\(T=0.21\) and \(T=0.32\)) display a clear structured signal.

The most prominent feature is a positive residual over approximately \(0.05<q<0.15~\mathrm{\AA^{-1}}\), followed by a negative residual over approximately \(0.16<q<0.30~\mathrm{\AA^{-1}}\). The magnitude of this structure increases substantially between the lowest and highest transmission settings.

The existence of these residuals demonstrates that the cysteine--water difference cannot be fully described by a detector-wide scale and offset correction. However, residual structure alone does not establish a sample-dependent effect because similar features could, in principle, arise from imperfect run-to-run reproducibility.

For this reason, the corresponding water--water controls must also be considered.

\subsection{Control-subtracted residuals}

Figure~\ref{fig:deep0p5M_cys_minus_control} shows the control-subtracted residuals, \(\Delta R(q)\), obtained after removal of the corresponding water--water controls.

The subtraction of the water controls does not eliminate the structured residual observed at higher transmission. Instead, the principal features remain clearly visible.

The lowest-transmission dataset remains comparatively weak, while the datasets at \(T=0.21\) and \(T=0.32\) retain a pronounced positive contribution over approximately \(0.05<q<0.15~\mathrm{\AA^{-1}}\) and a negative contribution over approximately \(0.16<q<0.30~\mathrm{\AA^{-1}}\).

The persistence of this structure after subtraction of the corresponding water controls indicates that it cannot be explained solely by the residual reproducibility limits observed within the water measurements.

\subsection{Integrated residual observables}

To quantify the transmission dependence of the residual structure, two integrated observables were defined,

\[
A_{+}
=
\int_{0.05}^{0.15}
\Delta R(q)\,dq
\]

and

\[
A_{-}
=
\int_{0.16}^{0.30}
\Delta R(q)\,dq.
\]

These quantities characterize the positive and negative lobes of the control-subtracted residual, respectively. The chosen integration windows correspond to the regions where the residual structure is most reproducible and least affected by low-\(q\) geometric uncertainties.

The resulting integrated observables are summarized in Table~\ref{tab:deep0p5M_delta_residuals}.

\begin{table}[htbp]
\centering
\caption{
Integrated control-subtracted residual observables for the 0.5 M cysteine dataset.
}
\label{tab:deep0p5M_delta_residuals}
\begin{tabular}{ccc}
\hline
Transmission & \(A_{+}\) & \(A_{-}\) \\
\hline
0.0099 & $-7.1\times10^{-4}$ & $-9.3\times10^{-4}$ \\
0.21   & $ 2.74\times10^{-3}$ & $-3.36\times10^{-3}$ \\
0.32   & $ 4.13\times10^{-3}$ & $-3.05\times10^{-3}$ \\
\hline
\end{tabular}
\end{table}

The lowest-transmission dataset exhibits only weak integrated residuals. In contrast, both observables increase substantially in magnitude at higher transmission. The positive lobe increases by approximately a factor of six between \(T=0.0099\) and \(T=0.21\), while remaining large at \(T=0.32\). A corresponding increase is observed for the negative lobe.

These integrated quantities provide a compact representation of the transmission dependence already visible in Fig.~\ref{fig:deep0p5M_cys_minus_control}. The residual signal is therefore not only apparent as a visual feature of the SAXS profiles but can also be quantified through simple scalar observables.

\subsection{Statistical significance of the integrated residuals}

The statistical robustness of the integrated observables was evaluated from the distribution of independent train-pair residual observables. For each transmission, the variability of the integrated residual observables was estimated directly from the train-pair ensemble.

For the two highest-transmission datasets, the integrated residual observables exceed their estimated statistical uncertainties by more than an order of magnitude. The corresponding significance levels range from approximately \(12\sigma\) to \(40\sigma\), confirming that the observed residual structure cannot be attributed to ordinary statistical fluctuations.

These values should not be interpreted as evidence for a specific physical mechanism. Rather, they demonstrate that the residual SAXS structure observed at higher transmission is statistically robust within the train-pair ensemble. The quoted significance levels reflect the statistical uncertainty estimated from the train-pair ensemble and do not include possible systematic uncertainties beyond those captured by the matched water controls.

\subsection{Similarity between the two higher-transmission datasets}

The two strongest residuals, corresponding to \(T=0.21\) and \(T=0.32\), were compared directly over the interval \(0.05<q<0.30~\mathrm{\AA^{-1}}\).

The resulting Pearson correlation coefficient is \(\rho=0.786\), indicating substantial similarity between the two residual shapes.

A best-fit linear scaling analysis shows that the \(T=0.32\) residual is larger in magnitude than the \(T=0.21\) residual but retains the same overall sign-changing structure. The two curves are therefore not identical, yet they appear to belong to the same family of residual profiles.

Taken together, the results presented in this section establish the principal experimental observation of this work: a reproducible, control-subtracted residual SAXS structure is present in the 0.5 M cysteine dataset and becomes increasingly pronounced with increasing transmission.

\section{Statistical robustness}

The existence of a structured residual signal does not by itself establish statistical robustness. A critical question is whether the observed residuals emerge progressively through averaging of many independent train pairs or whether they are dominated by a small subset of atypical events.

To address this question, the control-subtracted residuals were re-evaluated using progressively larger subsets of train pairs. In addition, block-averaging statistics were used to quantify the statistical stability of the integrated residual observables.

\begin{figure*}[htbp]
\centering
\includegraphics[width=\textwidth]{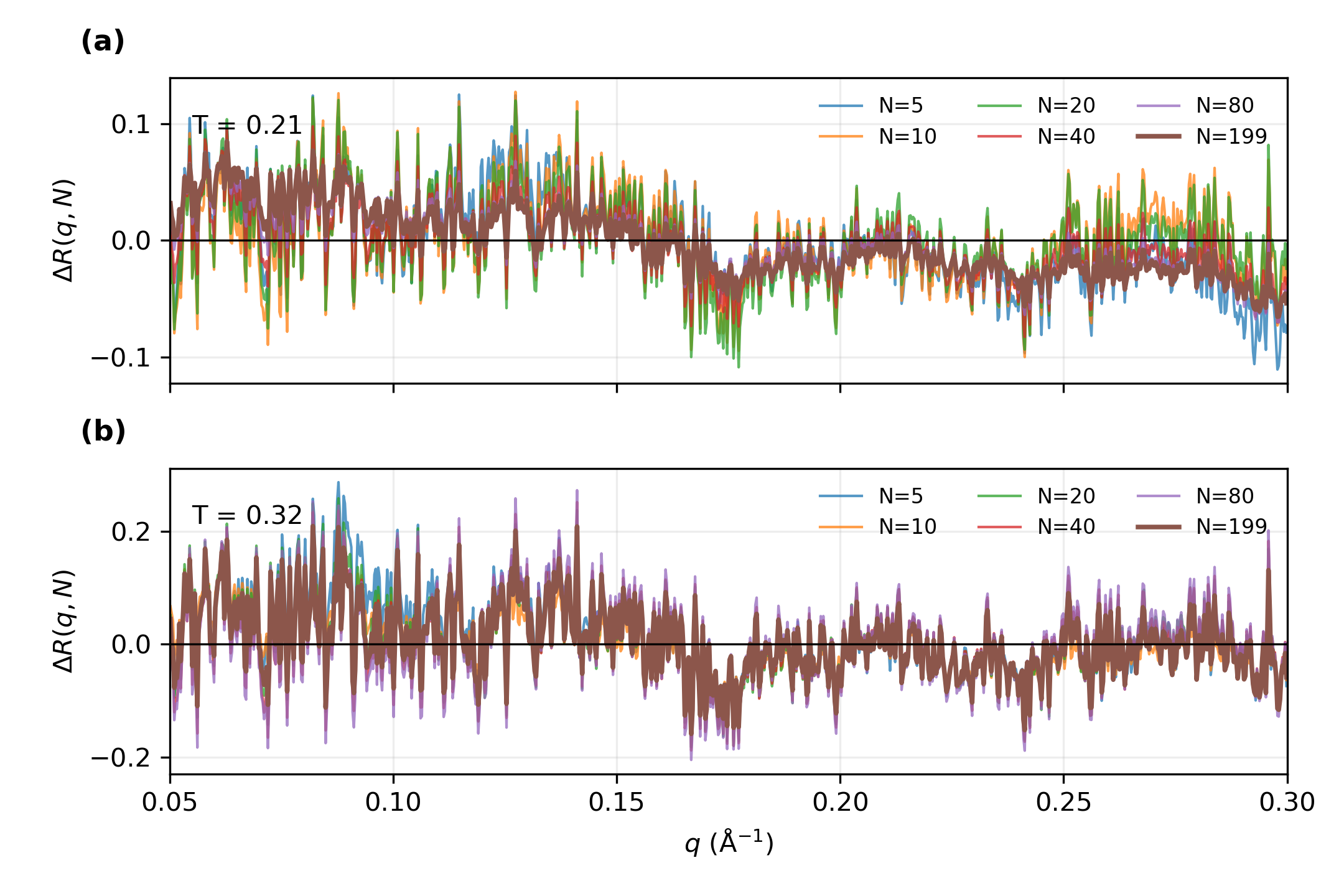}
\caption{
Convergence of the control-subtracted residual profiles as the number of train pairs increases. (a) \(T=0.21\). Curves obtained from progressively larger train-pair subsets approach the final \(N=200\) result while retaining the same overall sign-changing structure. (b) \(T=0.32\). The characteristic positive and negative residual lobes remain visible throughout the averaging process and progressively stabilize as additional train pairs are included.
}
\label{fig:convergence}
\end{figure*}
\begin{figure*}[htbp]
\centering
\includegraphics[width=\textwidth]{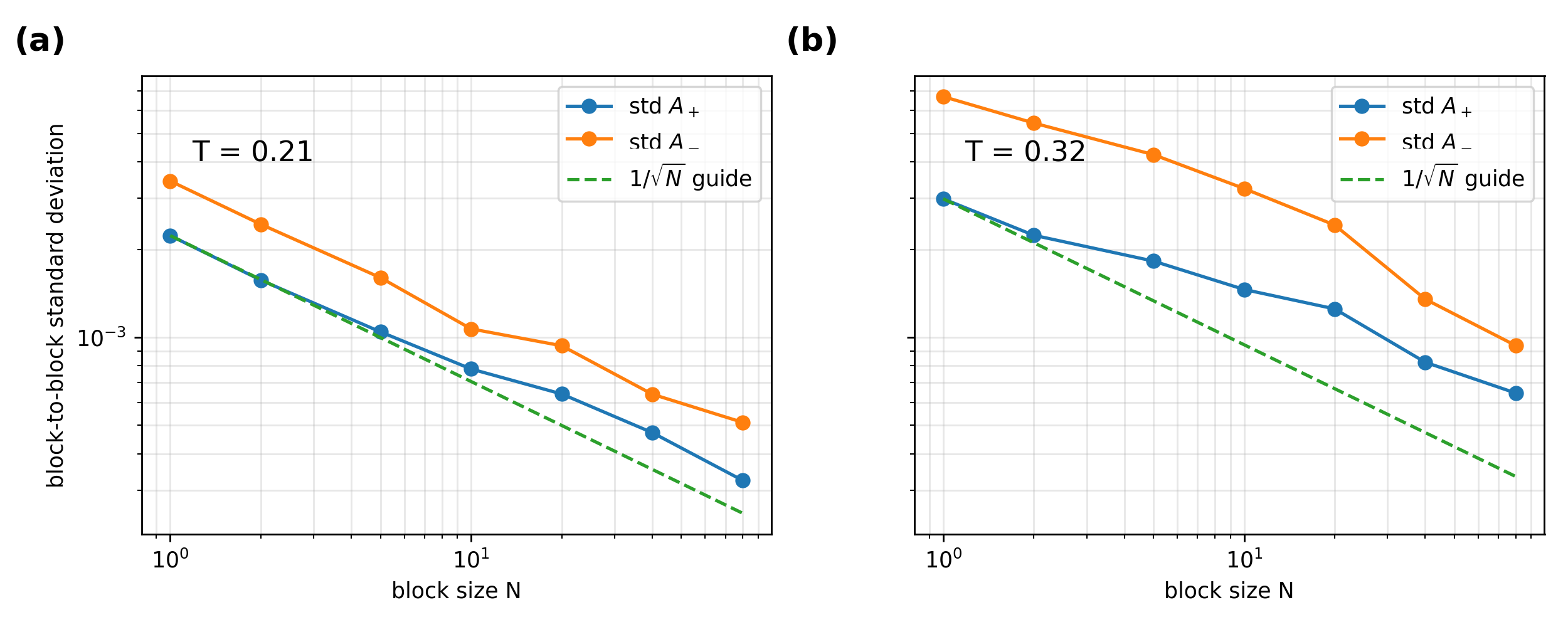}
\caption{
Block-averaging statistics of the integrated residual observables as a function of block size. (a) \(T=0.21\). (b) \(T=0.32\). For each block size \(N\), the train-pair residual observables were averaged within independent non-overlapping blocks, and the standard deviation of the resulting block averages was computed. The dashed curves indicate the expected \(1/\sqrt{N}\) scaling. In both datasets, the block-to-block variability decreases systematically with increasing \(N\), consistent with averaging statistically independent train-pair observables.
}
\label{fig:blockaveraging}
\end{figure*}

\subsection{Convergence of the residual profiles}

For each transmission, progressively larger subsets containing \(N=\{5,10,20,40,80,120,160,200\}\) train pairs were selected from the complete dataset and used to construct control-subtracted residual profiles.

Figure~\ref{fig:convergence} shows the resulting residual curves for the two transmission settings exhibiting the strongest signal.

For both transmission settings, the residual profiles progressively approach the final result as additional train pairs are accumulated. Increasing the number of train pairs primarily reduces statistical fluctuations while preserving the underlying structure.

No evidence is observed for the residual signal being generated by a small number of isolated train pairs. Instead, the residual emerges progressively and reproducibly as independent train pairs are accumulated.

\subsection{Block-averaging statistics}

A more direct assessment of experimental reproducibility was obtained using block averaging. For each block size \(N\), the train-pair residual observables were averaged within independent consecutive blocks, and the standard deviation across blocks was calculated.

Figure~\ref{fig:blockaveraging} summarizes the resulting block-to-block variability. The variability decreases systematically as the block size increases. No plateau is observed over the investigated range of \(N\), indicating that the residual observables remain dominated by statistical averaging rather than by an irreducible systematic floor.
For both transmission settings, the observed block-to-block variability is well approximated by
\[
\sigma(N)\propto\frac{1}{\sqrt{N}},
\]

as expected for statistically independent measurements.

\subsection{Implications for the residual signal}

The convergence and block-averaging analyses provide information that extends beyond the statistical significance of the residual itself. The control-subtracted residual observables exhibit approximately the expected \(1/\sqrt{N}\) reduction in block-to-block variability over the investigated range of block sizes. At the same time, the residual profile converges toward a stable shape as additional train pairs are accumulated.

This behaviour contrasts with earlier analyses of the raw cysteine--water SAXS difference signal within the present dataset, which showed substantially slower convergence than expected from ideal statistical averaging. The raw contrast is dominated by large correlated contributions arising from detector-wide intensity variations, normalization uncertainties, jet fluctuations, and other experimental common modes. Under these conditions, increasing the number of measurements alone does not guarantee a corresponding gain in sensitivity.

The present results demonstrate that train-level ON--OFF reconstruction, followed by scale-plus-offset correction and subtraction of transmission-matched water controls, reduces the dominant common-mode contributions and isolates a residual observable that approaches the statistical behaviour expected for an ensemble average of many train pairs. 

Once these correlated components are removed, the block-to-block variability decreases approximately as $1/\sqrt{N}$, allowing the high repetition rate of the European XFEL to be translated directly into statistical sensitivity.

This observation has important implications for future XFEL experiments aimed at detecting extremely weak signals. The principal advantage of high-repetition-rate operation is not simply the availability of a large number of pulses, but the possibility of constructing train-resolved observables for which common-mode fluctuations can be suppressed and statistical averaging can recover its expected behaviour. The present analysis therefore provides an experimental demonstration that weak residual scattering signals can emerge from large XFEL datasets when suitable controls and train-resolved analysis procedures are employed.

The data therefore behave as expected for a reproducible ensemble-averaged signal and illustrate the conditions under which high-repetition-rate XFEL operation can provide a genuine gain in sensitivity.

\section{Discussion}

The present analysis demonstrates the existence of a reproducible residual SAXS signal in the 0.5 M cysteine dataset after removal of train-resolved scale-plus-offset contributions and subtraction of transmission-matched water--water controls. The observation is therefore not explained by global intensity normalization effects alone. Even after independent optimization of the scale and offset parameters for every train pair, a structured residual remains.

The residual exhibits a reproducible momentum-transfer dependence. After subtraction of the matched water controls, the datasets acquired at \(T=0.21\) and \(T=0.32\) display the same qualitative pattern, consisting of a positive contribution over approximately \(0.05<q<0.15~\mathrm{\AA^{-1}}\) followed by a negative contribution over approximately \(0.16<q<0.30~\mathrm{\AA^{-1}}\). The similarity between these residual profiles, together with their positive Pearson correlation coefficient, indicates that the observed structure is not random. In addition, the signal amplitude increases substantially between \(T=0.0099\) and the two higher-transmission datasets, establishing a clear dependence on the incident transmission setting. The present data do not allow the functional form of this dependence to be determined. However, because the signal evolves systematically across the transmission settings, the observations are consistent with a beam-intensity-dependent contribution.

The residual also exceeds the variability measured in the water controls themselves. Furthermore, convergence and block-averaging analyses show that the signal becomes progressively better defined as independent train pairs are accumulated, while the variability of the integrated observables decreases approximately as expected from \(1/\sqrt{N}\) averaging. The observed residual is therefore not generated by a small number of anomalous train pairs but reflects a statistically robust ensemble property of the measurements.

An important aspect of the present analysis is that the observed residual structure remained qualitatively unchanged under multiple independent validation tests performed during development of the analysis workflow, including alternative normalization procedures, detector-level consistency checks, independent reprocessing paths, and verification of the train classification and averaging strategies. While no analysis can exclude every possible systematic effect, these tests substantially reduce the likelihood that the residual signal originates from a specific data-processing artifact.

At the same time, the present data do not uniquely identify the microscopic origin of the effect. The observed residual could arise from structural changes within the cysteine solution, modifications of the solvent environment, beam-induced density fluctuations, radiolysis products, hydrodynamic perturbations of the liquid jet, cavitation-related processes, or a combination of several mechanisms. The sign-changing shape of the residual profile suggests that the effect cannot be described as a simple overall increase or decrease in scattering intensity, but instead corresponds to a redistribution of scattering weight across momentum transfer.

The present study should therefore be regarded as an observational rather than a mechanistic investigation. In particular, no controlled concentration dependence can presently be established. Although measurements were also acquired for a 1 M cysteine solution, the absence of a complete set of transmission-matched water controls acquired in an alternating water--cysteine measurement sequence prevents a reliable comparison with the 0.5 M dataset. Similarly, the intermediate transmission dataset at \(T=0.046\) cannot presently be used to constrain the transmission dependence because the acquisition sequence was interrupted and experimental equivalence cannot be demonstrated with the same confidence as for the remaining datasets. The difficulties encountered in the 1 M dataset highlight the importance of alternating water and sample measurements when targeting weak XFEL solution-scattering signals. Even when residual structures are observed, their interpretation remains ambiguous unless corresponding water controls are available under closely comparable experimental conditions. The success of the 0.5 M analysis suggests that such acquisition strategies can substantially improve confidence that weak residual signals originate from the sample rather than from slow experimental drifts.

The principal outcome of this work is therefore the establishment of a reproducible residual SAXS signature that survives stringent internal controls and statistical validation. Future experiments employing fully alternated water and cysteine measurements, denser transmission sampling, and improved control statistics should allow the physical origin of this signal to be determined.

\section{Conclusions}

We have performed a train-resolved SAXS analysis of aqueous L-cysteine solutions measured at the European XFEL. By fitting each cysteine--water train pair with an independent scale-plus-offset model and subsequently subtracting transmission-matched water--water controls, we isolated residual scattering contributions that cannot be explained by global intensity scaling, additive offsets, or the residual reproducibility limits observed within the water dataset.

The 0.5 M measurements reveal a reproducible control-subtracted residual SAXS structure exhibiting a characteristic sign-changing dependence on momentum transfer. The residual is weak at the lowest transmission investigated and becomes stronger at higher transmission settings.

The two highest-transmission datasets display closely related residual shapes despite being obtained from independent measurements. The corresponding residual profiles exhibit a Pearson correlation coefficient of 0.786. Statistical analyses demonstrate that the observed residual represents a reproducible ensemble property of the train-pair dataset.

These measurements establish the existence of a reproducible transmission-dependent residual SAXS contribution in the 0.5 M cysteine dataset that survives both scale-plus-offset subtraction and matched water controls. The physical origin of this signal remains unresolved, and the current data do not uniquely distinguish between possible structural, chemical, hydrodynamic, or radiolytic mechanisms.

Beyond the observation of a reproducible residual SAXS signal, the present work establishes that train-resolved scale-plus-offset correction combined with transmission-matched water controls can define weak residual observables whose block-to-block variability decreases approximately as expected from \(1/\sqrt{N}\) averaging.

This result demonstrates that the unique high-repetition-rate capability of the European XFEL can be effectively converted into statistical sensitivity when suitable train-resolved observables and matched controls are employed. Weak residual signals that remain hidden in conventional averaged comparisons can therefore emerge after train-level ON--OFF reconstruction and accumulation of large train-pair ensembles. Similar statistical considerations are likely to be relevant for other XFEL measurements targeting intrinsically weak signals, including nonlinear X-ray processes.

\begin{acknowledgments}

The authors acknowledge European XFEL in Schenefeld, Germany, for the provision of X-ray free-electron laser beamtime under QUAST proposal 10917 (``High-repetition-rate SAXS screening of fluence-dependent X-ray--induced responses in liquids'').

C.S. acknowledges financial support from the Spanish Ministry of Science, Innovation and Universities through project PID2023-152154NB-C21. M.~C. acknowledges support from the European Research Council (ERC) Advanced Grant CHIRAX (No.~101095012). This work was also funded by the regional government of Madrid, Spain, through the Tecnolog\'ias 2024 program with the project MATRIX-CM (TEC-2024/TEC-85), the Ayudas a Proyectos Sin\'ergicos de I+D through the ACXIOM-CM project (SYG-2024/TEC-815), and the Ayudas de Atracci\'on de Talento Investigador C\'esar Nombela (Grant No.~2023-T2/ECO-28965). It was also supported by Grant PID2024-160991NA-I00 funded by MICIU/AEI/10.13039/501100011033 within the framework of the Generaci\'on de Conocimiento program.

The authors declare that they have no conflicts of interest.

The data that support the findings of this study are available from the corresponding author upon reasonable request.

\end{acknowledgments}

\bibliography{references}

\end{document}